\documentclass[twocolumn,showpacs,preprintnumbers,amsmath,amssymb,showkeys,nofootinbib]{revtex4}

\usepackage{graphicx}
\usepackage{dcolumn}
\usepackage{bm}
\newcommand{\bqa}{\begin{eqnarray}}
\newcommand{\eqa}{\end{eqnarray}}
\newcommand{\be}{\begin{equation}}
\newcommand{\ee}{\end{equation}}

\begin{document}


\title{$\Upsilon(4S,5S)\to\Upsilon(1S)\eta$ transitions in the rescattering model and the new BaBar measurement}

\author{Ce Meng$~^{(a)}$ and Kuang-Ta Chao$~^{(a,b)}$}
\affiliation{ {\footnotesize (a)~Department of Physics and State Key
Laboratory of Nuclear Physics and Technology, Peking University,
 Beijing 100871, China}\\
{\footnotesize (b)~Center for High Energy Physics, Peking
University, Beijing 100871, China}}


\begin{abstract}

The $\eta$ transitions of $\Upsilon(4S,5S)$ into $\Upsilon(1S,2S)$
are studied in the rescattering model by considering the final state
interactions above the $B\bar B$ threshold. The width of the $\eta$
transition of $\Upsilon(4S)$ into $\Upsilon(1S)$ is found to be
larger than that of the dipion transition, and the ratio of
$\Gamma(\Upsilon(4S)\to\Upsilon(1S)\eta)$ to
$\Gamma(\Upsilon(4S)\to\Upsilon(1S)\pi^+\pi^-)$ is predicted to be
$R_4=1.8\mbox{-}4.5$, which is about two orders of magnitude larger
than the expectation of the conventional hadronic transition theory,
and is supported by the new BaBar measurement. The widths of the
$\eta$ transitions of $\Upsilon(5S)$ are found to be sensitive to
the coupling constants $g_{\Upsilon(5S)B^{(*)}B^{(*)}}$ due to a
large cancelation between contributions from the $B\bar{B}$,
$B^*\bar{B}+c.c.$, and $B^*\bar{B}^*$ channels, and only a rough
estimate $\Gamma(\Upsilon(5S)\to\Upsilon(1S,2S)\eta)=10\mbox{-}200$
KeV can be given. The widths of the $\eta'$ transitions of
$\Upsilon(4S,5S)$ are also discussed, and they could be much smaller
than that of the corresponding $\eta$ transitions mainly due to the
tiny phase space.

\end{abstract}


\maketitle

\section{Introduction}

Hadronic transitions of heavy quarkonia are important for
understanding both the heavy quarkonium dynamics and the formation
of light hadrons. Because heavy quarkonium is expected to be compact
and non-relativistic, at least for the lower-lying states, QCD
multiple expansion (QCDME) approach (for recent reviews see,
e.g.~\cite{Kuang06,voloshin07}) can be used in the analysis of these
transitions. However, the justification of QCDME scenario becomes
problematic for higher-exited heavy quarkonia. Particularly, when
the excited state lies above the open flavor thresholds, the
coupled-channel effects may change the QCDME scenario markedly and
add new mechanisms to the analysis of its hadronic transitions.

In a previous paper~\cite{Meng-Y5S}, we use the final state
rescattering model~\cite{Cheng05_ReSC} to study the dipion
transitions of $\Upsilon(5S)$ and $\Upsilon(4S)$. In this model, the
$\Upsilon(5S/4S)$ first decays to $B^{(*)}\bar{B}^{(*)}$, and then
the $B$ meson pair turns into a lower $\Upsilon$ state and two pions
through exchange of another $B^{(*)}$ meson. We find that there is a
huge difference, which is about a factor of 200-600 in
magnitude~\cite{Meng-Y5S}, between the partial widths of dipion
transitions of $\Upsilon(5S)$ and $\Upsilon(4S)$. This result is
consistent with the measurement of the Belle
Collaboration~\cite{Belle07-5SmS}. The coupled-channel effects (or
the meson-loop effects) in the transitions
$\Upsilon(nS)\to\Upsilon(mS)\pi\pi$, where $n=2,3,4,5$ and $m<n$,
are also studied by Simonov et.
al.~\cite{Simonov0708-nSmS,Simonov08-5SmS}. In a recent calculation
for the dipion transitions of $\Upsilon(5S)$~\cite{Simonov08-5SmS},
their result confirms ours~\cite{Meng-Y5S} at the quantitative
level.

The above results indicate that it might be unnecessary to introduce
exotic interpretations for the $\Upsilon(5S)$ resonance such as the
$Y_b$ state~\cite{hou} to account for the experimental
data~\cite{Belle07-5SmS} if the rescattering model~\cite{Meng-Y5S}
can be proved efficient enough. Therefore, it is very useful to
study other features of the final state rescattering mechanism. One
evident feature of the rescattering mechanism is the strong
energy-dependence of the decay rates, which can induce significant
shifts of 10-20 MeV of the observed resonance peaks in
$\Upsilon(5S)\to\Upsilon(mS)\pi^+\pi^-$ relative to that of
$\Upsilon(5S)\to B^{(*)}\bar{B}^{(*)}$~\cite{Meng-PeakShift}.

Another important feature is that some of the power counting rules
in the QCDME approach may fail in the rescattering model. For
example, in the QCDME approach, the dipion transition of heavy
quarkonium can achieve through the E1-E1 (electric-dipole)
transition, whereas the $\eta$ transition is dominated by the E1-M2
(magnetic quadrupole) transition, which is associated with the
spin-flip effects of the heavy quarks, due to the $\eta$ quantum
number being $J^{PC}=0^{-+}$~\cite{Kuang06,voloshin07}. Therefore,
in the QCDME approach the $\eta$ transition is expected to be
strongly suppressed relative to the corresponding dipion transition,
whereas there is no such suppression in the rescattering model. In
fact, within the framework of the QCDME approach,
Kuang~\cite{Kuang06} predicted the ratios
\be\label{Rn:definition}
R_n=\frac{\Gamma(\Upsilon(nS)\to\Upsilon(1S)\eta)}{\Gamma(\Upsilon(nS)\to\Upsilon(1S)\pi^+\pi^-)}\sim10^{-2}\mbox{-}10^{-3}
\ee
for $n=2,3$, which are roughly in agreement with the new
measurements by the CLEO Collaboration~\cite{CLEO08-2S3S-1Seta}:
\bqa\label{R23:experiment} R_2 &=&
1.1^{+0.5}_{-0.4}\times10^{-3},\nonumber\\
R_3 &<& 7\times10^{-3}. \eqa
However, recently the preliminary result reported by the BaBar
Collaboration~\cite{Babar-4S1S-eta} indicates that for the
$\Upsilon(4S)$ the ratio
\be\label{R4:experimental}
R_4=\frac{\Gamma(\Upsilon(4S)\to\Upsilon(1S)\eta)}{\Gamma(\Upsilon(4S)\to\Upsilon(1S)\pi^+\pi^-)}=2.41\pm0.40\pm0.12,
\ee
which is larger than that for the $\Upsilon(2S,3S)$ in
(\ref{R23:experiment}) by two orders of magnitude or more. This is
another puzzling problem for hadronic transitions of heavy
quarkonium aside from the $\Upsilon(5S)$ dipion
transitions~\cite{Belle07-5SmS}. Here, again, a possible and natural
interpretation for this anomalously large difference between $R_4$
and $R_n$($n<4$) is that the rescattering mechanism is dominant in
the hadronic transitions of $\Upsilon(4S)$, since it lies above the
$B\bar B$ threshold, whereas $\Upsilon(nS)~(n<4)$ are below the open
bottom threshold and hence described by the conventional hadronic
transition theory.

In this paper, we will clarify whether the ratio $R_4$ can be as
large as (\ref{R4:experimental}) in the rescattering model, and give
some predictions for the $\eta$ transitions of $\Upsilon(5S)$ state.
We will first introduce the rescattering model and the notation of
$\eta-\eta'$ mixing in Sec.II. Then, we will numerically analyze the
$\eta$ as well as $\eta^{\prime}$ transitions of $\Upsilon(4S,5S)$
in turn in Sec.III. A summary will be given in the last section.

\section{The model}

\begin{figure}[t]
\begin{center}
\vspace{0cm}
 \hspace*{0cm}
\scalebox{0.5}{\includegraphics[width=16cm,height=18cm]{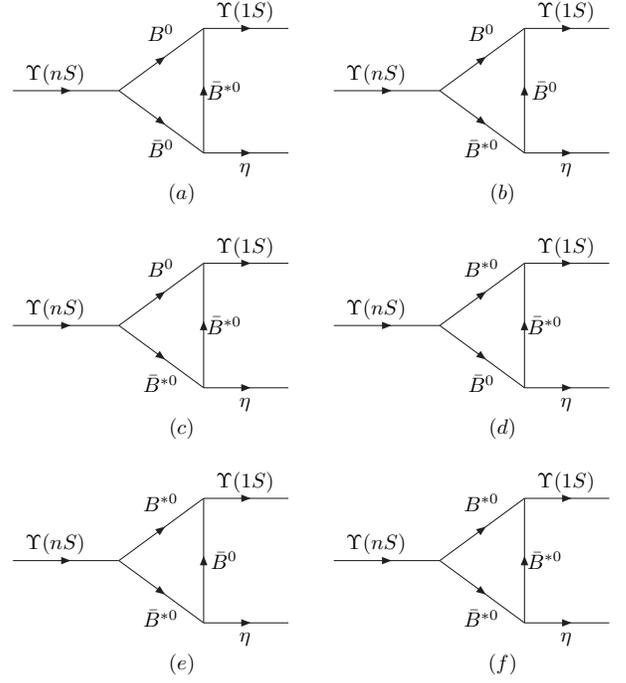}}
\end{center}
\vspace{0cm}\caption{The diagrams for $\Upsilon(nS)\to
B^{(*)0}\bar{B}^{(*)0}\to\Upsilon(1S)\eta$. Other diagrams can be
obtained by charge conjugation transformation
$B\leftrightarrow\bar{B}$ and isospin transformation
$B^0\leftrightarrow B^+$ and $\bar{B}^0\leftrightarrow
B^-$.}\label{Fig:Y-Yeta}
\end{figure}

In the rescattering model, the transitions
$\Upsilon(4S,5S)\to\Upsilon(1S)\eta$ can arise from scattering of
the intermediate state $B_{(s)}^{(*)}\bar{B}_{(s)}^{(*)}$ by
exchange of another $B_{(s)}^{(*)}$ meson. The typical diagrams for
the $B^{(*)}\bar{B}^{(*)}$ channels are shown in
Fig.~\ref{Fig:Y-Yeta}, and the other ones can be related to those in
Fig.~\ref{Fig:Y-Yeta} by the charge conjugation transformation
$B\leftrightarrow\bar{B}$ and isospin transformation
$B^0\leftrightarrow B^+$ and $\bar{B}^0\leftrightarrow B^-$.
Therefore, the amplitudes of Fig.~\ref{Fig:Y-Yeta}(a,b,c,d,e,f)
should be multiplied by a factor of 4, respectively. As for
$B_s^{(*)}\bar{B}_s^{(*)}$ channels, the typical diagrams are the
same as those in Fig.~\ref{Fig:Y-Yeta}, but the multiplied factor
should be 2.

To evaluate the amplitudes, we need the following effective
Lagrangians~\cite{Meng-Y5S,HMChEL}:
\begin{subequations} \label{effective-Lagrangians}
\begin{eqnarray}
\mathcal{L}_{\Upsilon BB}&=& g_{\Upsilon
BB}\Upsilon_\mu(\partial^\mu
B{B}^{\dagger}-B\partial^\mu {B}^{\dagger}),\label{L-YBB}\\
\mathcal{L}_{\Upsilon B^*B}&=& \!\frac{g_{\Upsilon\!
B^*\!B}}{m_{\Upsilon}}\varepsilon^{\mu\nu\alpha\beta}\partial_\mu
\!\Upsilon_\nu
\!\nonumber\\
&& \times(B^*_\alpha\overleftrightarrow{\partial}_\beta
{B}^{\dagger}\!\! - \!\!
B\overleftrightarrow{\partial}_\beta{B}^{*\dagger}_\alpha\!),\label{L-YB*B}\\
\mathcal{L}_{\Upsilon B^*B^*}&=& g_{\Upsilon B^* B^*} (
-\Upsilon^\mu
B^{*\nu}\overleftrightarrow{\partial}_\mu {B}_\nu^{*\dagger} \nonumber\\
&&+ \Upsilon^\mu B^{*\nu}\partial_\nu{B}^{*\dagger}_{\mu} -
\Upsilon_\mu\partial_\nu B^{*\mu}
{B}^{*\nu\dagger}),\label{L-YB*B*}\\
\mathcal{L}_{B^*B\eta}&=& ig_{B^*B\eta}B^*_{\mu}\partial^\mu\eta{B}^{\dagger},\label{L-B*Beta}\\
\mathcal{L}_{B^*B^*\eta}&=&
i\frac{g_{B^*B^*\eta}}{m_{B^*}}\varepsilon^{\mu\nu\alpha\beta}\partial_{\mu}B^*_{\nu}{B^*}^{\dagger}_{\alpha}
\partial_\beta\eta,\label{L-B*B*eta}
\end{eqnarray}
\end{subequations}
where
$\overleftrightarrow{\partial}=\overrightarrow{\partial}-\overleftarrow{\partial}$.
In the heavy quark limit, the coupling constants in
(\ref{effective-Lagrangians}) can be related to each other by heavy
quark spin symmetry as: \bqa g_{\Upsilon BB}&=&g_{\Upsilon
B^*B}=g_{\Upsilon B^*B^*}\,\label{HQS:g-YBB}\\
g_{B^*B\eta}&=&g_{B^*B^*\eta}.\label{HQS:g-SBB} \eqa Particularly,
the coupling constants for $\Upsilon(4S)$ and $\Upsilon(5S)$ can be
determined by the observed values of their partial decay widths to
the bottom meson pairs.

All the coupling constants will be determined below. However, it is
necessary to emphasize here that the determinations do not account
for the off-shell effects of the exchanged $B^{(*)}$ mesons, of
which the virtualities can not be ignored. Such effects can be
compensated by introducing, e.g., the monopole~\cite{Cheng05_ReSC}
form factors for off-shell vertexes. Let $q$ denote the momentum
transferred and $m_i$ the mass of exchanged meson, the form factor
can be written as
\begin{eqnarray}\label{formfactor1}
\mathcal{F}(m_{i},q^2)=\frac{(\Lambda+m_i)^{2}-m_{i}^2
}{(\Lambda+m_i)^{2}-q^{2}}.
\end{eqnarray}
For comparison, we will use the same cutoff $\Lambda=660$ MeV as the
one used in the numerical analysis of
$\Upsilon(4S,5S)\to\Upsilon(1S)\pi^+\pi^-$~\cite{Meng-Y5S}.

In the lightest pseudoscalar-meson nonet of the $SU(3)$ quark model,
there are two iso-scalar components, which can be written in the
octet-singlet basis as
\bqa \eta_8&=&\sqrt{\frac{1}{6}}(u\bar{u}+d\bar{d}-2s\bar{s}),\nonumber\\
\eta_0&=&\sqrt{\frac{1}{3}}(u\bar{u}+d\bar{d}+s\bar{s}),
\nonumber\eqa
where $\eta_8$ is one of the Goldstone bosons in the octet
representation of the chiral symmetry. If the intrinsic glue
component is negligible, the physical wave-functions of $\eta$ and
$\eta'$ can the be written as
\bqa |\eta\rangle&=\mbox{cos}\theta_{P}|\eta_8\rangle-\mbox{sin}\theta_{P}|\eta_0\rangle,\nonumber\\
|\eta'\rangle&=\mbox{sin}\theta_{P}|\eta_8\rangle+\mbox{cos}\theta_{P}|\eta_0\rangle,\label{eta:mixing}
\eqa
where the mixing angle $\theta_P$ has been determined in many places
in the literature and the value is in the range from $-13^\circ$ to
$-22^\circ$ (see, e.g., Ref.~\cite{Thomas-etaeta'-mixing}). We will
choose a moderate value
\be \theta_P=-17^\circ, \nonumber\ee
in our numerical analysis.

As the first step, we will treat the $\eta$ as a pure $\eta_8$
state, and leave the mixing effect to be considered in the following
section. An evident advantage of this treatment is that one can
relate the coupling constant $g_{B^*B\eta}$ to  $g_{D^*D\pi}$ using
heavy quark flavor symmetry and chiral symmetry~\cite{HMChEL}:
\bqa\label{g:BBeta}
g_{B^*B\eta}&=&-\frac{1}{2}g_{B_s^*B_s\eta}=\frac{1}{\sqrt6}\frac{\sqrt{m_Bm_{B^*}}}{\sqrt{m_Dm_{D^*}}}g_{D^*D\pi}\nonumber\\
&=&\frac{\sqrt2}{\sqrt3}\frac{\sqrt{m_Bm_{B^*}}}{f_\pi}g, \eqa
where $f_\pi=131$ MeV is the $\pi$ decay constant, and the coupling
constant $g\approx0.6$ is determined by the measurement of the decay
width of $D^{*+}\to D^0\pi^+$~\cite{PDG06}.

In the rescattering $\Upsilon(4S,5S)\to
B\bar{B}\to\Upsilon(1S)\eta$, the intermediate process
$\Upsilon(4S,5S)\to B\bar{B}$ can take place in a real or virtual
way, which corresponds to the imaginary part or the real part of the
amplitude, respectively. If the $\Upsilon(nS)$ state lies above the
$B^{(*)}\bar{B}^{(*)}$ threshold, the absorptive part (imaginary
part) of the amplitude arising from Fig.~\ref{Fig:Y-Yeta} can be
evaluated by the Cutkosky rule. For the process $\Upsilon(nS)\to
B^{(*)}(p_1)+\bar{B}^{(*)}\to \Upsilon(mS)+\eta$, the absorptive
part of the amplitude reads
\bqa\label{Abs:CutRule}
\textbf{Abs}_i&=&\frac{|\vec{p}_1|}{32\pi^2m_{\Upsilon(nS)}}\int
d\Omega
\mathcal{A}_i(\Upsilon(nS)\to B^{(*)}\bar{B}^{(*)})\nonumber\\
&&\times \mathcal{C}_i(B^{(*)}\bar{B}^{(*)}\to \Upsilon(mS)\eta),
\eqa
where $i=(a,b,c,d,e,f)$, and $d\Omega$ and $\vec{p}_1$ denote the
solid angle of the on-shell $B^{(*)}\bar{B}^{(*)}$ system and the
3-momentum of the on-shell $B^{(*)}$ meson in the rest frame of
$\Upsilon(nS)$, respectively.

The evaluation of the real part of the amplitude is difficult to be
achieved, and will bring large uncertainties inevitably.
Fortunately, for the transitions
$\Upsilon(4S,5S)\to\Upsilon(1S)\eta$, the contributions from the
real part are expected to be small, because the masses of
$\Upsilon(4S,5S)$ are not very close to the open flavor thresholds
as those of $X(3872)$~\cite{Meng07_X3872_ReSC} and
$Z(4430)$~\cite{Meng07_Z4430_ReSC}. In the previous
paper~\cite{Meng-Y5S}, we have roughly estimated the contributions
to the dipion transitions of $\Upsilon(4S,5S)$ from the real parts
through the dispersion relation and found that these contributions
are negligible for $\Upsilon(5S)$ and somewhat comparable to those
from the imaginary parts for $\Upsilon(4S)$. The same argument
should be also valid here for the $\eta$ transitions of
$\Upsilon(4S,5S)$.
As in \cite{Meng-Y5S}, we will neglect the contribution from the
real part and use (\ref{Abs:CutRule}) to determine the full
amplitude in the calculations. This scheme is efficient enough to
serve our aims.

In the absorptive part, which corresponds to the real rescattering
process, the intermediate states $B^{(*)}\bar{B}^{(*)}$ are
on-shell. Similar to the case of the dipion transitions of
$\Upsilon(4S,5S)$~\cite{Meng-Y5S}, the amplitude in
(\ref{Abs:CutRule}) is proportional to $|\vec{p}_1|^3$. This very
fact results in both the large difference between the dipion
transition rates of $\Upsilon(5S)$ and
$\Upsilon(4S)$~\cite{Meng-Y5S} and the markedly peak shift effect in
$\Upsilon(5S)\to\Upsilon(mS)\pi^+\pi^-$~\cite{Meng-PeakShift}. One
can generally expect the similar effects to emerge in the
corresponding $\eta$ transitions.

\section{Numerical Results and Discussions}

Since the contribution from the imaginary part of the re-scattering
amplitude corresponds to the real decay process $\Upsilon(nS)\to
B^{(*)}\bar{B}^{(*)}$, the coupling constants
$g_{\Upsilon(nS)B^{(*)}B^{(*)}}$ should be determined by the
measured values of the decay widths of $\Upsilon(4S,5S)\to
B^{(*)}\bar{B}^{(*)}$~\cite{PDG06}, and the results are given
by~\cite{Meng-Y5S}
\bqa g_{\Upsilon(4S)BB}&=& 24,\label{g:Upsilon(4S)BB}\\
 g_{\Upsilon(5S)BB}&=& 2.5,\label{g:Upsilon(5S)BB}\\
 g_{\Upsilon(5S)B^*B}&=& 1.4\pm0.3,\label{g:Upsilon(5S)B*B}\\
 g_{\Upsilon(5S)B^*B^*}&=& 2.5\pm0.4.\label{g:Upsilon(5S)B*B*}\eqa
The value of $g_{\Upsilon(4S)BB}$ in (\ref{g:Upsilon(4S)BB}) is
typical, and is comparable to the estimation using the vector meson
dominance model~\cite{Meng07_X3872_ReSC} for $g_{\Upsilon(1S)BB}$:
\be g_{\Upsilon(1S)BB}\thickapprox
\frac{m_{\Upsilon(1S)}}{f_{\Upsilon(1S)}}\sim
15,\label{g:UpsilonBB:VMD}\ee
where the decay constant $f_{\Upsilon(1S)}$ can be determined by the
leptonic width of $\Upsilon(1S)$. However, the values determined
from the $\Upsilon(5S)$ data in
(\ref{g:Upsilon(5S)BB})-(\ref{g:Upsilon(5S)B*B*}) are small. This
may be partly due to the fact that as a higher-excited $b\bar{b}$
state, the wave function of $\Upsilon(5S)$ has a complicated node
structure (with four nodes), and the coupling constants will be
small if the $p$-values $|\vec{p_1}|$ of $B^{(*)}\bar{B}^{(*)}$
channels (1060-1270 MeV) are close to those corresponding to the
zeros in the decay amplitude. The symmetry relation in
(\ref{HQS:g-YBB}) can also be violated by the same reason. This fact
has been confirmed by a specific calculation~\cite{Simonov08-5SBB},
recently.

Following Ref.~\cite{Meng-Y5S}, for the other coupling constants
$g_{\Upsilon(mS)B^{(*)}B^{(*)}}$ ($m<5$), we assume that the
symmetry relations in (\ref{HQS:g-YBB}) hold, and they are equal to
each other, which is implied by comparison between
(\ref{g:Upsilon(4S)BB}) and (\ref{g:UpsilonBB:VMD}).

Since the rescattering amplitude is proportional to $|\vec{p_1}|^3$,
one can expected that the contributions from $B_s\bar{B}_s$ channels
are generally much smaller than those from $B\bar{B}$ channels,
although there is an enhancement factor of 2 in the coupling
constant $g_{B_s^*B_s\eta}$ in (\ref{g:BBeta}). Therefore, here we
only use the central value of the decay widths of $\Upsilon(5S)\to
B_s^{(*)}\bar{B}_s^{(*)}$~\cite{Belle08-5sBs} to determine  the
coupling constant $g_{\Upsilon(5S)B_s^{(*)}B_s^{(*)}}$:
\bqa
 g_{\Upsilon(5S)B_sB_s}&=& 1.4,\nonumber\\
 g_{\Upsilon(5S)B_s^*B_s}&=& 2.0,\nonumber\\
 g_{\Upsilon(5S)B_s^*B_s^*}&=& 7.5.\label{g:Upsilon(5S)Bs*Bs*}\eqa
Here, the coupling constant $g_{\Upsilon(5S)B_s^*B_s^*}$ is larger
than the others and those in
(\ref{g:Upsilon(5S)BB}-\ref{g:Upsilon(5S)B*B*}). This can be
understood by the fact that the $p$-value of $B_s^{*}\bar{B}_s^{*}$
channel $|\vec{p_1}|\approx0.48$ GeV is small, which makes the
amplitude away from the zero points
sufficiently~\cite{Simonov08-5SBB}. This can also serve as evidence
in favor of the usual bottomonium interpretation of $\Upsilon(5S)$
in addition to its usual leptonic decay width~\cite{PDG06}.

\subsection{$\Upsilon(4S)\to\Upsilon(1S)\eta$}

Only Fig.~\ref{Fig:Y-Yeta}(a) is allowed for the real rescattering
process $\Upsilon(4S)\to B\bar{B}\to\Upsilon(1S)\eta$. For a pure
$\eta_8$ component, the result reads
\be
\Gamma(\Upsilon(4S)\to\Upsilon(1S)\eta)=2.92~\mbox{KeV}.\label{Gamma4S1Seta}\ee
Together with the prediction for the width
$\Gamma(\Upsilon(4S)\to\Upsilon(1S)\pi^+\pi^-)=(1.47\pm0.03)$ KeV in
the same model~\cite{Meng-Y5S}, we can get the ratio
\be R_4\simeq2.0,\label{R4:theory}\ee
which is consistent with the experimental
measurement~\cite{Babar-4S1S-eta} in (\ref{R4:experimental}).

Although the absolute value of the width in (\ref{Gamma4S1Seta})
suffers from large uncertainties due to the cutoff $\Lambda$, the
real part contamination, and the coupling constants
$g_{\Upsilon(1S)B^{*}B}$, the situation for the ratio $R_4$ in
(\ref{R4:theory}) should be better, since many of the uncertainties
canceled out in the ratio. On the other hand, the ratio is indeed
sensitive to the description of the production of $\pi^+\pi^-$. In
Ref.~\cite{Meng-Y5S}, we assume that the scalar resonance ($\sigma$,
$f_0(980)$...) contributions are dominant in the dipion production
and estimate the coupling constant $g_{\sigma BB}$ through symmetry
and re-scaling analysis. The value of $g_{\sigma BB}$ used by us is
lager than the one~\cite{Liu08-g:sigma} deduced from linear
representation of chiral symmetry~\cite{Bardeen03-g:sigma} by 20$\%$
in magnitude. The later value~\cite{Liu08-g:sigma} will enhance the
ratio in (\ref{R4:theory}) by a factor of 1.5.

Another large uncertainty of $R_4$ comes from the mixing between
$\eta_8$ and $\eta_0$ in (\ref{eta:mixing}). While the mixing angle
$\theta_P$ is rather well determined, there is no reliable
information for the coupling constant $g_{B^*B\eta_0}$.  As a
tentative assumption, we choose the value of $g_{B^*B\eta_0}$ in the
range from zero to the value of $g_{B^*B\eta_8}$ determined in
(\ref{g:BBeta}), and then the results are given by
\bqa R_4& = &1.8\mbox{-}3.1,\label{R4:mixing}\\
\Gamma(\Upsilon(4S)\to\Upsilon(1S)\eta')&=&0.1\mbox{-}0.3~\mbox{KeV}.\label{Gamma4S1Seta'}\eqa
Here, the width $\Gamma(\Upsilon(4S)\to\Upsilon(1S)\eta')$ is very
small mainly due to the tiny phase space.

To sum up, we find the ratio to be $R_4=1.8\mbox{-}4.5$ in the
rescattering model, which agrees with the experimental
measurement~\cite{Babar-4S1S-eta} in (\ref{R4:experimental}). This
can serve as another evidence for the dominant role of the
rescattering mechanism  in the hadronic transitions of
$\Upsilon(4S,5S)$, which lie above the open bottom threshold.

\subsection{$\Upsilon(5S)\to\Upsilon(1S/2S)\eta$}

To study the real rescattering effects in the transitions
$\Upsilon(5S)\to\Upsilon(1S/2S)\eta$, one needs to evaluate the
imaginary part of the amplitudes for all the diagrams in
Fig.~\ref{Fig:Y-Yeta} and for both $B^{(*)}\bar{B}^{(*)}$ channels
and $B^{(*)}_s\bar{B}^{(*)}_s$ channels. The contributions from the
$B^{(*)}_s\bar{B}^{(*)}_s$ channels are very small as one can see
later. Thus, in Tab.~\ref{Tab:5S1Seta}, we only list the
contributions from the $B\bar{B}$, $B^*\bar{B}+c.c.$ and
$B^*\bar{B}^*$ channels respectively, and totally.

We use the central values in
(\ref{g:Upsilon(5S)BB}-\ref{g:Upsilon(5S)B*B*}) to evaluate the
transition width obtained from each single channel. The results
shown in Tab.~\ref{Tab:5S1Seta} are all about 100 KeV, and are much
greater than the width of $\Upsilon(4S)\to\Upsilon(1S)\eta$ in
(\ref{Gamma4S1Seta}). The reason is just the same as the large
difference between the dipion transition widths of $\Upsilon(5S)$
and $\Upsilon(4S)$, namely the $\Upsilon(5S)$ has much larger
$|\vec{p}|$ values than $\Upsilon(4S)$.

However,, there is a large cancelation between these three channels.
As a result, the total widths of these transitions, which are listed
in the last line of Tab.~\ref{Tab:5S1Seta}, are very small. This
cancelation makes the widths to be very sensitive to the coupling
constants determined in
(\ref{g:Upsilon(5S)BB}-\ref{g:Upsilon(5S)B*B*}), which can be seen
through the large error bars in Tab.~\ref{Tab:5S1Seta}. If we choose
all the coupling constants $g_{\Upsilon(5S)B^{(*)}B^{(*)}}=2.5$, the
calculated widths for the $5\to1$ and $5\to2$ transitions will be
145 and 83 KeV, respectively. So, these two widths can not be
determined accurately and we can only give loose estimates for them:
\bqa \Gamma(\Upsilon(5S)\to\Upsilon(1S)\eta)& = &20\mbox{-}150~\mbox{KeV},\label{Gamma5S1Seta}\\
\Gamma(\Upsilon(5S)\to\Upsilon(2S)\eta)&=&10\mbox{-}100~\mbox{KeV}.\label{Gamma5S2Seta}\eqa

\begin{center}\begin{table}
\caption{Transition widths of $\Upsilon(5S)\to
B^{(*)}\bar{B}^{(*)}\to\Upsilon(mS)\eta$ in units of KeV. The error
bars come from those of $g_{\Upsilon(5S)B^*B}$ and
$g_{\Upsilon(5S)B^*B^*}$.}
\begin{tabular}{ccc}
\hline
Channel & $\Upsilon(5S)\to\Upsilon(1S)\eta$ & ~~$\Upsilon(5S)\to\Upsilon(2S)\eta$  \\
\hline  $B\bar{B}$ &
~~80 & ~~78 \\
  $B^*\bar{B}+c.c.$ & ~~70 & ~~59 \\
  $B^*\bar{B}^*$ & ~~141 & ~~172 \\
  total &
~~$30^{+22+24}_{-17-17}$ & ~~$9^{+13+17}_{-7-8}$\\
\hline \label{Tab:5S1Seta}\end{tabular}
\end{table}\end{center}

As for the contributions from the $B^{(*)}_s\bar{B}^{(*)}_s$
channels, they are very small as we have mentioned. Choosing the
values of the coupling constants in ({\ref{g:Upsilon(5S)Bs*Bs*}}),
the total decay widths from these channels are only about 1 KeV.
Needless to say, there is also a large cancelation between these
channels.   Moreover, the width for an individual channel is only
about 10 KeV, which is much smaller than those from
$B^{(*)}\bar{B}^{(*)}$ channels.

Similar to the case of $\Upsilon(4S)\to\Upsilon(1S)\eta$, the mixing
between $\eta$ and $\eta'$ can cause additional uncertainties of
50$\%$ in magnitude to the decay widths in (\ref{Gamma5S1Seta}) and
(\ref{Gamma5S2Seta}). The width of the transition
$\Upsilon(5S)\to\Upsilon(1S)\eta'$ is about 10-40 KeV due to the
mixing.

\section{Summary and discussion}

In summary, we study the effects of long-distance final state
interactions on the $\eta$ transitions of $\Upsilon(4S,5S)$ in the
rescattering model. We find that the width of the $\eta$ transition
of $\Upsilon(4S)$ to $\Upsilon(1S)$ is larger than that of the
dipion transition, and the ratio of the former to the latter is
predicted to be $R_4=1.8\mbox{-}4.5$, which is consistent with the
experimental data~\cite{Babar-4S1S-eta}. This result, together with
those in Ref.~\cite{Meng-Y5S}, indicate that the real rescattering
mechanism can be dominant in the hadronic transitions of the higher
$\Upsilon$-states that lie above the $B\bar{B}$ threshold.

Estimations for the virtue rescattering effects can be roughly made
through the same procedure by using the dispersion relation as that
suggested in Ref.~\cite{Meng-Y5S}. The contributions are 3-6 KeV and
1.1-1.5 KeV to the widths, respectively, of the $\eta$ and dipion
transitions of $\Upsilon(4S)$ to $\Upsilon(1S)$, while the ratio
$R_4$ in (\ref{R4:theory}) will be enhanced by a factor of 1.2-1.5.
In addition, the virtual rescattering contributions are found to be
about (1-2)$\times10^{-2}$ KeV and (5-6)$\times10^{-4}$ KeV to the
widths $\Gamma(\Upsilon(nS)\to\Upsilon(1S)\eta)$ for $n=3$ and 2,
respectively. The former is larger than but roughly consistent with
the upper limit of the width measured by CLEO
Collaboration~\cite{CLEO08-2S3S-1Seta}, while the latter is smaller
than the measurement~\cite{CLEO08-2S3S-1Seta} by an order of
magnitude, which indicates that the QCDME mechanism may be dominant
in the transition $\Upsilon(2S)\to\Upsilon(1S)\eta$ since
$\Upsilon(2S)$ is far below the $B\bar B$ threshold. Note that the
absolute values of these transitions are sensitive to the values for
the coupling constants, e.g., $g_{\Upsilon(1S)B^{*}B}$. If the value
in (\ref{g:UpsilonBB:VMD}) is used instead of that in
(\ref{g:Upsilon(4S)BB}) for the $g_{\Upsilon(1S)B^{*}B}$, the
absolute transition widths of $\Upsilon(4S)\to\Upsilon(1S)\eta$ and
$\Upsilon(4S)\to\Upsilon(1S)\pi\pi$ as well as
$\Upsilon(3S)\to\Upsilon(1S)\eta$ and
$\Upsilon(2S)\to\Upsilon(1S)\eta$ will decrease by almost a factor
of 3, while the ratio $R_4$ remain unchanged, and thus all
predictions for $\Upsilon(nS)\to\Upsilon(1S)\eta (n=4,3,2)$ can
become consistent with observed data. Another point is that the
above estimations for the virtue rescattering effects depend on the
cutoff parameter $\Delta$ (see Ref.~\cite{Meng-Y5S} for details),
and we have chosen $\Delta=m_{B^*}-m_{B}$ in our calculations as in
Ref.~\cite{Meng-Y5S}.

As for the $\eta$ transitions of $\Upsilon(5S)$, the widths are very
sensitive to the coupling constants $g_{\Upsilon(5S)B^{(*)}B^{(*)}}$
because there is a large cancelation between the contributions from
$B\bar{B}$, $B^*\bar{B}+c.c.$ and $B^*\bar{B}^*$ channels. Thus we
can only give very loose estimations
$\Gamma(\Upsilon(5S)\to\Upsilon(1S,2S)\eta)=10\mbox{-}200$ KeV.

Besides, the widths of the $\eta'$ transitions of $\Upsilon(4S,5S)$
are generally expected to be much smaller than those of the
corresponding $\eta$ transitions due to the tiny phase space, but we
need to have a better understanding for the couplings of the
flavor-singlet $\eta_0$ to $B^{(*)}B^{(*)}$ meson pairs before we
can draw a definite conclusion on the $\eta'$ transitions.

In conclusion, the observed anomalously large $\eta$ transition rate
of $\Upsilon(4S)$ to $\Upsilon(1S)$ might be explained in the
rescattering model above the open bottom threshold, despite of large
uncertainties in chosen parameters.

~~~~~~~~~~~~~~~~~~

$Note.$ When this manuscript was completed, a paper on the
transitions of $\Upsilon(nS)\to\Upsilon(1S)\eta$ ($n=2,3,4,5$)
appeared~\cite{Simonov08-nS-1Seta}. The predicted width of
$\Upsilon(4S)\to\Upsilon(1S)\eta$ is in agreement with ours and the
predicted width of $\Upsilon(5S)\to\Upsilon(1S)\eta$ is close to the
lower limit given by us, but without large error bars as ours.
However, the predicted width of
$\Upsilon(3S)\to\Upsilon(1S)\eta$~\cite{Simonov08-nS-1Seta} is
greater than the experimental upper limit in (\ref{R23:experiment})
by a factor of 200-500.

~~~~~~~~~~~~~~~~~~

\begin{acknowledgments}
We wish to thank H.Q. Zheng for helpful discussions. This work was
supported in part by the National Natural Science Foundation of
China (No 10675003, No 10721063).
\end{acknowledgments}

\bibliography{apssamp}

\end{document}